\documentclass[twocolumn,showpacs,amsmath,amssymb,prd,nofootinbib]{revtex4}
\usepackage{epsfig}
\def\sss{\scriptscriptstyle}
\def\^#1{^{\sss #1}}
\def\_#1{_{\sss #1}}
\def\beq{\begin{equation}}
\def\eeqno#1{\label{#1}\end{equation}}

\def\cmss{{\rm cm~s^{-2}}}

\def\az{a\_{0}}

\def\l0{\ell\_{0}}

\def\rar{\rightarrow}

\def\l{\lambda}

\def\m{\mu}
\def\n{\nu}

\def\a{\alpha}

\def\gN{g\_N}

\begin{document}
\title{MOND impact on and of the recently updated mass-discrepancy-acceleration relation}
\author{Mordehai Milgrom}
\affiliation{Department of Particle Physics and Astrophysics, Weizmann Institute}

\begin{abstract}
McGaugh et al. (2016) have recently used their extensive SPARC sample of disc galaxies to update the well-known mass-discrepancy-acceleration relation (MDAR): a tight correlation between the local, dynamical accelerations, $g$, deduced from the rotation curves at different radii, and the local baryonic, Newtonian gravitational accelerations, $\gN$. This MDAR is one of the major predicted ``MOND laws''. It is anything but a newly discovered relation, as it has been plotted and studied time and again (starting with Sanders 1990), with ever increasing quantity and quality of data. The important updated analysis of McGaugh et al., like its precedents, bears crucial ramifications for the observed dynamical anomalies in disc galaxies, and, in particular, on their resolution by the MOND paradigm. These results, indeed, constitute a triumph for MOND.
However, unlike previous analyses of the MDAR, McGaugh et al. have chosen to obfuscate the MOND roots of their analysis, and its connection with, and implications for, this paradigm.
For example, they find that their data are very well fit by a certain, single-parameter function, apparently
picked {\it ex machina}, of the form $g=\gN\nu(\gN/g_\dag)$, with $\nu(y)=(1-e^{-\sqrt{y}})^{-1}$, and $g_\dag\approx 1.2\times 10^{-8}\cmss$. In fact, this
$\n(y)$ tightly follows the dictates of MOND: a single acceleration parameter, $\nu(y\gg 1)\approx 1$, and $\nu(y\ll 1)\approx y^{-1/2}$, from scale invariance of the deep-MOND limit. This very same $\n(y)$
has been proposed and used repeatedly in the past in the context of MOND. No other possible origin for such a function is known. It was already known to reproduce rotation curves correctly, so was {\it a priori} assured success here as well.
That $g_\dag$ plays the role, and has the well-established value, of the MOND acceleration constant, $\az$, is also suppressed.
The present paper corrects these oversights -- bringing to light the deep connections with MOND, suppressed by McGaugh et al. It also gives due credit to previous works, and discusses some important, but less known, aspects of this MOND relation.
\end{abstract}
\pacs{04.50.Kd, 95.35.+d}
\maketitle

\section{\label{introduction} Introduction}
MOND \cite{milgrom83} contends that the dynamical anomalies in galactic systems are due to departure from the standard dynamics at low accelerations, not to the presence of dark matter. The basic tenets of this paradigm are as follows: a. Dynamics is characterized by a single ``boundary acceleration constant'', $\az$. b. For system accelerations $\gg\az$, standard dynamics obtains (Newtonian correspondence). c. For accelerations much below $\az$ (the deep-MOND limit) MOND greatly departs from standard dynamics, with MOND dynamics being space-time scale invariant.
MOND is reviewed, e.g., in Refs. \cite{fm12,milgrom14c}.
\par
One of the major predictions of MOND -- a MOND law -- put forth in the very first MOND paper \cite{milgrom83} [Eq.(2) there], is that the measured ``dynamical'' acceleration, $g$, -- e.g., that measured from the rotation curve as $V^2(R)/R$ -- will be found to be tightly correlated with the `baryonic', Newtonian acceleration, $\gN$, at the same position $g\approx f(\gN)$.\footnote{In a class of so called modified-inertia MOND theories \cite{milgrom94}, this is predicted to be a functional relation, with no scatter, when applied to rotation curves of disc galaxies, as in most studies. In ``modified gravity'' MOND theories this relation is expected to be tight, but not functional (e.g., Ref. \cite{brada95}).}
Moreover, MOND has dictated a very restricted form of $f$: Since $\az$ is the only acceleration constant in the theory, we must have, on dimensional grounds, $g\approx \gN\n(\gN/\az)$. Newtonian correspondence dictates that $\n(y\gg 1)\approx 1$. Scale invariance in the deep-MOND limit (plus the normalization of $\az$) dictates that $\n(y\ll 1)\approx y^{-1/2}$.
\par
This fundamental MOND prediction has been put to the test many times in the past.
The first, to my knowledge, to test this from rotation curves of disc galaxies was Sanders \cite{sanders90}.

\par
The test consists of plotting any of several equivalent forms of $g$ vs. $\gN$ for disc, or late type galaxies: Either as $g/\gN$ vs.
$g$ \cite{sanders90,mcgaugh99,mcgaugh04}, or vs. $\gN$ \cite{mcgaugh04,tiret09,wu15,desmond16,janz16}.
Or, equivalently, one can plot $g$ vs. $\gN$ directly, as done in Refs. \cite{scarpa03,wu15}.
This progression of studies is driven by continual improvement in the quantity and quality of the data.
\par
The fact that $g\not =\gN$ constitutes the mass discrepancy, and $g/\gN$, or $g-\gN$, can be its quantitative measures. Hence the common name of this relation: the mass-discrepancy-acceleration-relation (MDAR).
\par
Refs. \cite{sanders90,mcgaugh99,mcgaugh04}, also plotted $g/\gN$ against size, to find that there is no correlation with size, but with acceleration, as MOND predicts.
\par
All those previous studies of the $g$ vs. $\gN$ relation acknowledge the roots in MOND, the direct MOND impetus to plot this relation, and the great relevance to MOND that this tight correlation has.
\section{The updated MDAR}
Now, McGaugh et al. \cite{mcgaugh16}, in the latest in this succession, have improved on previous analyses, by using their extended SPARC sample of disc galaxies.\footnote{Ref. \cite{desmond16} already used the SPARC sample for this purpose.} These improvements are described in Ref. \cite{mcgaugh16} and earlier references therein. Their analysis involves many more galaxies, with larger dynamic range of the various galaxy properties (luminosity, mass, size, surface brightness). Additionally, photometry in the
$3.6~\m$ is used to deduce the stellar mass distribution, as the luminosity in this spectral region is thought to be a better representative of the stellar mass, with relatively small variation in the mass-to-light conversion factor. \par
This, more advantageous, photometry was already used in Ref. \cite{gfb13} for MOND analysis of rotation curves, which underlies the MDAR, but for the much smaller THINGS sample of only twelve galaxies.
\par
Ref. \cite{mcgaugh16} plots $g$ vs. $\gN$ and finds, like previous studies, that they are tightly correlated.
They show that their data are very well represented by a function (picked for undisclosed reason, but see below) with a single acceleration-parameter, of the form $g=\gN\nu(\gN/g_\dag)$, with
\beq \nu(y)=(1-e^{-\sqrt{y}})^{-1}.\eeqno{i}
 They find the best fit value of $g_\dag\approx 1.2\times 10^{-8}\cmss$.
\par
Contrary to the previous studies of this very same relation (including by McGaugh himself), Ref. \cite{mcgaugh16} gives only a vague lip-service, in passing, to the influence of, and relevance of all this, to MOND.
\par
It should be clear, however, even to a superficial observer, that this whole analysis, including the choice of the fitting function (see bellow) constitute MOND analysis par excellence.
\par
The results of Ref. \cite{mcgaugh16}, are indeed, in all their details, a clear verification of the relevant MOND prediction.

\section{Choice of the fitting function}

It could appear that the choice of the very restricted form of $\n(y)$, invoked by Ref. \cite{mcgaugh16} -- like the whole idea -- resulted from some deep acumen.
In fact, however, this very same $\n(y)$  has been proposed and used repeatedly in the past (including by McGaugh himself), always as a MOND interpolation function. As such, it was taken from the start to follows the strict dictates of the basic MOND tenet: a. It involves a single acceleration parameter, called in Ref. \cite{mcgaugh16} $g_\dag$, but which is but a guise for the MOND acceleration $\az$. b. For large arguments, $\nu(y\gg 1)\approx 1$, dictated by Newtonian correspondence at large accelerations. c. $\nu(y\ll 1)\approx y^{-1/2}$, which follows from the scale invariance of the low-acceleration, deep-MOND limit.
\par
There is no other paradigm, certainly not the dark-matter paradigm, that dictates this form.
The results of Ref. \cite{mcgaugh16} should thus be views as a triumph for MOND.
\par
This function was introduced, and used for the first time, in Ref. \cite{ms08}
[designated there as $\bar\n\_{1/2}(y)$] as part of a MOND analysis of cluster lensing. It was later used
by McGaugh, in Ref. \cite{mcgaugh08}, for MOND analysis of the Milkey way rotation curve, and in the MOND review, Ref. \cite{fm12}, as their MOND function of choice for MOND rotation-curve fits. It was also discussed in Ref. \cite{milgrom12} in connection with weak lensing in MOND.
\par
So, after it had been well established that this $\nu(y)$ reproduces correctly the observed rotation curves from the baryon distribution, it is obvious {\it a priori} that it will fit well the MDAR, which is but a summary of many rotation curves.

It should be clear, though, that there is nothing very special in this particular function, and it should  not be given special significance. Many other and varied analytic forms of $\n(y)$ that
abide by the MOND dictates would fit the data just as well. For example, the choice
\beq \n(y)=[1+(1+4y^{-1})^{1/2}]/2,\eeqno{ii}
 which is also widely used in MOND rotation curve analysis, and which corresponds to the MOND $\m(x)=x/(1+x)$, gives a practically indistinguishable curve for the acceleration range studied.\footnote{This function, however, approaches 1 too slowly for $y\rar\infty$.}
\par
A quick inspection will show that, by and large, any smooth function, with the MOND asymptotes, that has no small parameters, and for which
$1.45\lesssim\n(1)\lesssim 1.75$, would fit the MDAR very well. Another set of such functions is
 \beq \nu(y)={\rm tgh}^{-\a}(y^{1/2\a}),\eeqno{iii}
  with $\a\approx 1.5-2$.
These functions differ from those of eqs.(\ref{i}-\ref{ii}) by a few percents at the most. $\a=1.75$ gives a particularly close match.
\subsection{The acceleration constant}
What Ref. \cite{mcgaugh16} call $g_\dag$ is but their guise for the MOND acceleration constant $\az$. It appears in several roles in the MDAR, as per the original MOND prediction (see section \ref{comments}). The value Ref. \cite{mcgaugh16} found, $\az=(1.20\pm 0.26)\times 10^{-8}\cmss$, practically coincides with the values obtained in several previous MOND rotation-curve analyses. For example, already Ref. \cite{milgrom88} found, with rather primitive rotation-curves data,  $\az=(1.3\pm 0.5)\times 10^{-8}\cmss$. With rather better data, Ref. \cite{bbs91} found $\az=(1.21\pm 0.27)\times 10^{-8}\cmss$, and McGaugh himself, in a test of the MOND prediction of the mass-asymptotic-speed relation, found $\az=(1.24\pm 0.14)\times 10^{-8}\cmss$  \cite{mcgaugh11}.

\section{\label{comments}General comments on the MDAR}
I take this opportunity to make some additional comments on the MDAR and its relevance to MOND.
\par
The MDAR is a summary of rotation-curves and baryon-distribution data for many galaxies. As any good summary it has its merits. But there are important aspects that it does not capture from the analysis of individual galaxies. Notably, rotation-curve analysis shows that in some instances there are pronounced features in the baryon distribution that are clearly reflected in the measured $V(R)$ -- as predicted by MOND, but not for dark matter. But this cannot be seen in a MDAR plot: Even a pronounced feature in $V(R)$ simply performs a back-and forth motion along the smooth MDAR. This is a loss of crucial information that only full analysis can show.
\par
More generally, there is much more information in the full MOND rotation-curve prediction than in the prediction of where the different points lie in the $g-\gN$ plane.\footnote{For example, a galaxy could have nearly constant $g$ and $\gN$ over a stretch of radii, corresponding to only one point on the MDAR.}
\par
It is interesting to note that the MOND acceleration, $\az$, enters the MDAR in three independent ways, and
the fact that they all agree with each other is no simple feat of MOND, and would be a tall order for the dark-matter paradigm (see, e.g., Ref. \cite{milgrom14c}):  $\az$ enters the predicted MDAR as the transition acceleration from the Newtonian to the deep-MOND regime. The other two appearances ocur as follows: As also noted in Ref. \cite{mcgaugh16} (without mentioning MOND), the low-acceleration part of the MDAR [where $g\approx (\az\gN)^{1/2}$] picks up two types of contributions: a. Contributions from the asymptotic parts of rotation curves, predicted by MOND, together with the MOND, baryonic-mass-asymptotic speed relation: For these points, $V=(MG\az)^{1/4}$; so clearly $g=V^2/R= (\az MG/R^2)^{1/2}=(\az\gN)^{1/2}$. This is the second appearance of $\az$. The other contribution comes from the bulk of galaxies that are wholly in the deep-MOND regime. There, $\az$ enters as setting the magnitude of the dynamical anomalies within the galaxies.

\section{Summary and discussion}
I have argued, and demonstrated, I believe, that:

a. The recent study of Ref. \cite{mcgaugh16} is an important advance over previous studies of the MDAR.

b. It involves no conceptually new insight, or a newly discovered relation, as one is given to understand. Its main merit is in the use of more and better data over previous studies of the same relation.

c. The notion of searching for such a relation over time, and the form of the analytic function used to describe it are wholly rooted in, and motivated by MOND, and by what had already been known about its performance in describing rotation curves and, indeed, the MDAR itself.

d. The updated results are in full agreement with the predictions of MOND.
\par
McGaugh et al. seem to try to justify their suppressing the role of MOND, by saying that one ``should be careful not to confuse data with theory''. However, in the first place, we are not speaking here of data -- the data itself is not even theirs -- we are speaking of regularities and laws that have been mined from the data. And, while it is true that to some level, theory and observed phenomena should be considered independently, this does not justify ignoring the role of the theory in predicting and pointing analysts to look for these laws. (In the present case, the law under discussion is not even a new discovery, as I have shown.)
\par
It hardly needs repeating that the role of physical theories is not only to explain known facts. Even more importantly, a good theory makes predictions, and directs experimentalists or observers to look for new phenomena, regularities, and interconnections, not previously known.
\par
This is exactly what MOND did in the case of the MDAR, which has not been discovered by serendipity or shrewdness. It was looked for and plotted because MOND predicted it.
\par
The same is true of the MOND mass-asymptotic-speed-relation, where MOND predicted zero scatter in a version of the Tully-Fisher relation when the total baryonic mass is plotted against the constant asymptotic rotational speed of a galaxy. (In particular, Ref. \cite{mb88} emphasized the imperative to include the gas mass in addition to the stellar mass.) It was only then that people started using this particular velocity measure in the Tully-Fisher relation (for the first time in Ref. \cite{sanders96}, who was testing the MOND prediction) in what has come to be known as the baryonic Tully-Fisher relation.

And the same is true of other MOND laws, including the full description of measured rotation curves with the ``MOND formula''.
\par
The asymptotic flatness of rotation curves is a counterexample, as its emergence preceded MOND -- and indeed it was the main inspiration for the creation of MOND,  as the measurement of the black-body spectrum was the impetus for quantum mechanics, or Kepler's and Galilei's laws for Newtonian dynamics, or Balmer's laws for the quantum theory of atoms.
\par
Taking the independence of phenomena and theory in the sense adopted by McGaugh et al. could have lead the LIGO experiment to pretend that they have discovered some new phenomenon of ``heaviness undulations'', suppressing the role of general relativity in instigating, and narrowly directing, the search and detection for this phenomenon.
Or it might have lead the LHC to ignore the role of the standard model of particle physics in the searching and finding of the Higgs boson.
\par
Yes, after the finding, one should not consider the case shut, and allow for the possibility that the phenomenon in question could result from other theories. But one should still emphasize the support that the finding lends to the theory that did predict it.

\end{document}